# A Friction Model of Fractal Rough Surfaces Accounting for Size Dependence at Nanoscale

X. M. Liang and G. F. Wang*

Department of Engineering Mechanics, SVL and MMML, Xi'an Jiaotong University, Xi'an 710049, China

* E-mail: wanggf@mail.xjtu.edu.cn

**Abstract**

Traditional laws of friction believe that the friction coefficient of two specific solids takes constant value. However, molecular simulations revealed that the friction coefficient of nanosized asperity depends strongly on contact size and asperity radius. Since contacting surfaces are always rough consisting of asperities varying dramatically in geometric size, a theoretical model is developed to predict the friction behavior of fractal rough surfaces in this work. The result of atomic-scale simulations of sphere-on-flat friction is summarized into a uniform expression. Then, the size dependent feature of friction at nanoscale is incorporated into the analysis of fractal rough surfaces. The obtained results display the dependence of friction coefficient on roughness, material properties and load. It is revealed that the friction coefficient decreases with increasing contact area or external load. This model gives a theoretical guideline for the prediction of friction coefficient and the design of friction pairs.

# 1. Introduction

Classical laws of friction believe that the maximal static friction force is proportional to the applied load and independent of the nominal contact area. In that case, the friction coefficient takes constant value independent of external load. However, some experiments [1-9] reported the dependence of friction coefficient on external load. Generally, the friction coefficient tends to decrease as the load applied increases for various friction pairs. Many researches dedicated to explaining this phenomenon from a theoretical perspective. Usually, theoretical friction models are established in a two-step pattern.

First, the contact of rough surfaces should be properly described. Multi-asperity contact models have attracted most attention due to their simplicity. By modeling the rough surface as a nominal flat plane with randomly distributed spherical asperities, Greenwood and Williamson [10] established a famous theoretical contact model. In light of this idea, a lot of asperity models have been brought up. Generally, these contact models predict an approximately linear [10-12] or sub-linear [1, 13] relationship between the real contact area and load. On the other hand, the fractal nature of rough surfaces has long been observed [14]. Accordingly, Majumdar and Tien [15] adopted the Weierstrass-Mandelbrot function to describe self-affine rough surfaces. Since then, many researches on fractal contact models have been conducted [16-18]. Besides, a prominent contact model was proposed by Persson [19, 20], which attributes the contact fraction to different magnifications of surface roughness.

In addition, properly describing the single-asperity friction behaviors is indispensable to the full understanding of rough surfaces friction. There are basically two different strategies being used to characterize the single-asperity friction. Some researchers thought that the local shear stress exists merely on real contact surfaces and takes constant value. With the aforementioned linear or sub-linear load-area relationship, a corresponding constant or decreasing friction coefficient with increasing external load would be predicted. Others applied the von Mises criterion on the stress field of a sliding sphere [21], and thought that the plastic deformation is the reason why an asperity failed [22, 23] or partly failed [24, 25] to support any additional tangential force. As the normal load increases, more and more asperities tend to deform plastically, which leads to a decreasing friction coefficient.

However, experiments by AFM and SFA demonstrated that the mean friction stress varies from 20 MPa to 860 MPa as the tip radius shrinks from 1 cm [26] to 140 nm [27]. Li and Kim measured the mean friction stress between a gold nano-asperity and a flat mica, and observed a clearly size-dependent friction behavior [28]. Nanoscale asperity friction experiments on LFM [29, 30] also verified that the friction force varies with normal load, which might be potentially attributed to the changing contact radii. Therefore, the mean friction stress is related to the contact size and asperity radius. Hurtado and Kim [31, 32] proposed a model regarding the evolution of partial slip as the nucleation and gliding of a circle dislocation. An obvious size-dependence of the mean tangential stress on contact area was presented. In a certain range of contact radius $r$, the mean tangential stress initially takes constant value and then drops as $r^{-1/2}$ [31]. Atomic simulations also showed that mean tangential stress decreases as $r^{-1/2}$ for commensurate contact surfaces with adhesion and $r^{-3/2}$ for incommensurate ones [33]. For commensurate sphere-on-flat friction with purely repulsive interfacial interaction, Sharp et al. [34] found that the friction coefficient drops as $r^{-2/3}$ in a certain range. Such scaling law was accounted for by Wang et al. [35] based on a semi-analytical method.

In this study, we present a fractal friction model of elastic-plastic contacting rough surfaces. For single-asperity, a general expression summarized from atomic-scale simulations is employed to account for the size-dependent friction behavior. By applying this expression to each asperity on fractal rough surfaces, the relationships between friction force, external load and the real contact area are subsequently obtained. Furthermore, an unambiguous dependence of friction coefficient on the contact area and external load is illustrated. Comparison between this model and some available experiment results show a good agreement.

## 2. Majumdar-Bhushan fractal contact model

First, the fractal contact model by Majumdar and Bhushan (MB model) is briefly reviewed. In the MB model [16], a surface profile is formed by the superposition of cosinusoidal waves of varying wavelengths. For a specific wavelength $l$, the undeformed asperity shape is given as

$$z(x) = W^{D-1} l^{2-D} \cos\left(\frac{\pi x}{l}\right), \quad \left(-\frac{l}{2} < x < \frac{l}{2}\right) \tag{1}$$

where $W$ is a characteristic length scale of surface, and $D$ is the fractal dimension. Both $W$ and $D$ can be determined by the power spectrum of the surface. The area of a contact spot can be estimated by $a = l^2$. Then the curvature radius $R$ of the asperity tip can be derived as

$$R = \left| \left( \frac{d^2 z}{dx^2} \bigg|_{x=0} \right)^{-1} \right| = \frac{a^{D/2}}{\pi^2 W^{D-1}} \quad (2)$$

The real contact area $A_r$ is only a small portion of the apparent contact area $A_a$ and consists of contact spots of various sizes. The sizes of contact spots are assumed to follow a power law distribution

$$n(a) = \frac{D}{2} \frac{a_l^{D/2}}{a^{D/2+1}} \quad (3)$$

where $a_l$ is the area of the largest contact spot. The real contact area can thus be calculated as

$$A_r = \int_0^{a_l} n(a) a \, da = \frac{D}{2-D} a_l \quad (4)$$

A critical contact area $a_c$ distinguishing whether asperity deforms elastically or plastically is determined by

$$a_c = \frac{W^2}{(B\phi/2)^{2/(D-1)}} \quad (5)$$

where $B$ is the ratio of the hardness $H$ to the yield stress $\sigma_y$, and $\phi$ is a dimensionless material parameter given by $\phi = \sigma_y/E^*$.

For a contact spot with its area larger than $a_c$, this asperity will deform purely elastically. The normal load acting on this asperity can be calculated using Hertzian theory and Eq. (2) as

$$P_e(a) = \frac{4\sqrt{\pi}}{3} E^* W^{D-1} a^{(3-D)/2} \quad (6)$$

Similarly, when the area of a contact spot is smaller than $a_c$, the asperity will undergo plastic deformation and the normal load is written as

$$P_p(a) = B\sigma_y a \tag{7}$$

Therefore, the total external load $P$ can be calculated as

$$P = \begin{cases} \int_{a_c}^{a_l} P_e(a) n(a) \mathrm{d}a + \int_0^{a_c} P_p(a) n(a) \mathrm{d}a, & \kappa > 1 \\ \int_0^{a_l} P_p(a) n(a) \mathrm{d}a, & \kappa \leq 1 \end{cases} \tag{8}$$

where $\kappa$ is the ratio of $a_l$ to $a_c$. When $\kappa$ is smaller than 1, the areas of all contact spots do not reach the critical contact area $a_c$, and all asperities deform plastically. Once the ratio $\kappa$ increases beyond 1, a portion of contact spots will undergo pure elastic deformation.

By substituting Eqs. (3)-(7) into Eq. (8), the dimensionless total external load is then obtained. For $D \neq 1.5$ and $\kappa > 1$, the external load is expressed as

$$\begin{aligned} P^* &= \frac{4\sqrt{\pi}}{3} \frac{D}{3-2D} \left( \frac{2-D}{D} \right)^{(3-D)/2} W^{*D-1} A_r^{*(3-D)/2} \\ &+ \left( \frac{2D}{2-D} + \frac{4\sqrt{\pi}}{3} \frac{D}{2D-3} \right) \left( \frac{2-D}{D} \right)^{D/2} (B\phi/2)^{(2D-3)/(D-1)} W^{*2-D} A_r^{*D/2} \end{aligned} \tag{9a}$$

And for $\kappa \leq 1$, the external load gives

$$P^* = B\phi A_r^* \tag{9b}$$

where

$$\begin{aligned} P^* &= \frac{P}{A_a E^*}, \quad W^* = \frac{W}{\sqrt{A_a}}, \quad A_r^* = \frac{A_r}{A_a} \\ \kappa &= a_l / a_c = A_r^* \Big/ \left( \frac{W^{*2}}{(B\phi/2)^{2/(D-1)}} \frac{D}{2-D} \right) \end{aligned} \tag{10}$$

Obviously, by combining the MB model and the size-dependent friction coefficient of a single asperity, the dependence of the maximal static friction force on the real contact area can be conducted accordingly.

## 3. Friction between rough surfaces

As a preparatory work, some previous researches on micro-sized asperity friction are summarized into a general expression first. Generally, most theoretical models [31, 32, 35] and simulations [33, 34, 36] predict either two or three regimes of size-dependent asperity friction.

For a repulsive sphere-on-flat contact, when the contact size is pretty small, all atoms move concurrently, and the mean friction coefficient $\mu$ equals to the constant local maximal friction coefficient $\alpha$, also known as the static friction coefficient for a single atom [37]. For mid-large contact size, a dislocation nucleates at the contact edge and rapidly sweeps throughout the contact region. The critical shear stress for dislocation nucleation will decay as the contact radius $r$ increases. As a result, $\mu$ decreases linearly as $r$ increases in double logarithmic coordinates, as shown in Fig. 1. As the contact size gets even large, $\mu$ rises again because the Peierls stress [38] will increase as $r$ increases. Only the first two regimes are considered here, and the third regime can be dealt with accordingly for further research.

For commensurate repulsive contact, the friction coefficient extracted from atomic scale simulations [34, 35] are shown in Fig. 1. These results suggest that the frictional behavior of a single asperity can be characterized by

$$\mu/\alpha = \left(1+\gamma c\frac{\alpha r^2}{Rd}\right)^{-2/3} \tag{11}$$

where $\gamma$ is a fitting parameter, $c$ equals to $4/[\pi(1-\nu)]$ with $\nu$ being the Possion's ratio, $d$ is the atomic separation, and $R$ represents the curvature radius of spherical asperity. For the considered case in Fig. 1, Eq. (11) shows quantitatively well agreement with simulation results for $\gamma = 0.4374$.

Based on the general form of the friction coefficient of a single asperity and the fractal contact model in Sec. 2, the friction behavior of rough surfaces can be evaluated. As a simplification, it is assumed that the normal load and the shear force are independent. For single-asperity friction, such assumption is supported by some researches [34, 35, 39, 40], in which the Hertzian relation still holds for asperity contact during sliding. As for rough surface friction, experiments [9] showed that the normal load remain approximately unchanged during the onset of friction, which means the coupling effect, if existed, could be negligible. Substituting Eq. (2) into Eq. (11), the asperity friction coefficient is expressed as

$$\mu(a) = \alpha\left(1+\frac{4\gamma\alpha}{d(1-\nu)}W^{D-1}a^{1-D/2}\right)^{-2/3} \tag{12}$$

We assume that the contacting rough surfaces remain at rest till all micro-asperities reach their maximal tangential force. Then the maximal static friction force $F$ is given by

$$F = \begin{cases} \int_{a_c}^{a_l} \mu(a)P_e(a)n(a)\mathrm{d}a + \int_0^{a_c} \mu(a)P_p(a)n(a)\mathrm{d}a, & \kappa > 1 \\ \int_0^{a_l} \mu(a)P_p(a)n(a)\mathrm{d}a, & \kappa \leq 1 \end{cases} \tag{13}$$

By substituting Eqs. (3)-(7) and Eq. (12) into Eq. (13), the maximal static friction force is then obtained. For $D \neq 1.5$ and $\kappa > 1$, the dimensionless maximal static friction force is expressed as

$$F^* = \frac{4\sqrt{\pi}}{3}\alpha \frac{D}{2D-3}\left(\frac{2-D}{D}\right)^{D/2}(B\phi/2)^{(2D-3)/(D-1)}W^{*2-D}A_r^{*D/2}f\left(-\frac{\eta^*W^*}{(B\phi/2)^{(2-D)/(D-1)}}\right)$$

$$-\frac{4\sqrt{\pi}}{3}\alpha \frac{D}{2D-3}\left(\frac{2-D}{D}\right)^{(3-D)/2}W^{*D-1}A_r^{*(3-D)/2}f\left(-\left(\frac{2-D}{D}\right)^{1-D/2}\eta^*W^{*D-1}A_r^{*1-D/2}\right)$$

$$+3\alpha\left(\frac{2-D}{D}\right)^{D/2-1}B\phi W^{*1-D}A_r^{*D/2}\eta^{*-1}\left[\left(1+\frac{\eta^*W^*}{(B\phi/2)^{(2-D)/(D-1)}}\right)^{1/3}-1\right]$$

(14a)

And for the case $\kappa \leq 1$, it gives

$$F^* = 3\alpha\left(\frac{2-D}{D}\right)^{D/2-1}B\phi W^{*1-D}A_r^{*D/2}\eta^{*-1}\left[\left(1+\left(\frac{2-D}{D}\right)^{1-D/2}\eta^*W^{*D-1}A_r^{*1-D/2}\right)^{1/3}-1\right]$$

(14b)

where

$$F^* = \frac{F}{A_a E^*}, \quad \eta^* = \eta\sqrt{A_a}, \quad \eta = 4\gamma\alpha/[d(1-\nu)]$$
$$f(z) = hypergeom\left(\frac{2}{3},\frac{3-2D}{2-D};\frac{5-3D}{2-D};z\right)$$

(15)

and *hypergeom*(*a*, *b*; *c*; *z*) is the hypergeometric function expressed as

$$hypergeom(a,b;c;z) = \sum_{n=0}^{\infty}\frac{(a)_n(b)_n}{(c)_n}\frac{z^n}{n!}$$
$$(q)_n = \begin{cases} 1, & n=0 \\ q(q+1)\cdots(q+n-1), & n>0 \end{cases}$$

(16)

With Eq. (9) and Eq. (14), the normalized external load $P^*$ and maximal static friction force $F^*$ are obtained.

## 4. Results and discussions

Four compound parameters are here to be considered, i.e. $\eta^*$, $B\phi$, $D$, and $W^*$. The first parameter $\eta^*$ indicates the influence of lattice structure, whereas $B\phi$, equaling $H/E^*$, is introduced to evaluate the influence of plastic deformation on the friction force. The effects of rough topography are accounted for through $D$ and $W^*$.

Fig. 2 displays the dependence of $F^*$ on the normalized real contact area $A_r^*$ with various $\eta^*$. The conventional prediction $\alpha P^*$ is also included for comparison. In our analysis, the friction force $F^*$ first increases along the same curve as $\alpha P^*$ does, then the increasing rate of $F^*$ will slow down as $A_r^*$ increases. Therefore, a decreasing ratio of $F^*$ to $P^*$ is predicted. A large value of $\eta^*$ tends to accelerate this decreasing process and vice versa. This could be easily conducted from Eq. (14). When the contact area is extremely small, $F^*$ becomes

$$F^* \approx \alpha B\phi A_r^* = \alpha P^* \tag{17}$$

Therefore, Eq. (17) is an asymptotic line for $F^*$. The enlargement of $\eta^*$ will make $F^*$ deviate from the asymptotic line as shown in Fig. 2.

For $B\phi$ within a common range, the variations of dimensionless friction force $F^*$ with respect to the normalized real contact area $A_r^*$ are shown in Fig. 3. Generally, $F^*$ increases monotonically as the contact fraction $A_r^*$ gets larger. When the contact fraction is small, the influence of plasticity is prominent, and surface with a larger $B\phi$ leads to a higher friction force. As the contact area increasing, the normalized friction force $F^*$ converges, and the influence of plastic deformation gradually fades out thereafter.

For the fractal dimension $D$ in the range of [1.28, 1.52], the friction-area curves are displayed in Fig. 4. It seems that the fractal dimension $D$ has little influence on the relation. While the friction force is evidently increased as $W^*$ gets large, as shown in Fig. 5.

The total friction coefficient $\mu_s$ of the contacting rough surfaces is defined as the ratio of the maximal static friction to the normal load. For various values of $\eta^*$, Fig. 6 demonstrates the variation of $\mu_s/\alpha$ with respect to the normalized real contact area $A_r^*$. For very small contact area, the friction coefficient of rough surfaces $\mu_s$ takes the same value as the local friction coefficient $\alpha$. As the contact area increases, $\mu_s$ drops at an increasing rate in the double logarithmic coordinates. Finally, this rate reaches an upper bound, and $\mu_s/\alpha$ yields a power law relationship with $A_r^*$. The enlargement of $\eta^*$ will speed up the decreasing process of $\mu_s$.

The total friction coefficient $\mu_s$ likewise changes with varying $B\phi$, as shown in Fig. 7. When $\kappa$ takes value less than or equal to unity, dividing Eq. (14b) by Eq. (9b) gives $\mu_s/\alpha$ as

$$\mu_s/\alpha = 3\left(\frac{2-D}{D}\right)^{D/2-1} W^{*1-D} A_r^{*D/2-1} \eta^{*-1} \left[\left(1+\left(\frac{2-D}{D}\right)^{1-D/2} \eta^* W^{*D-1} A_r^{*1-D/2}\right)^{1/3} - 1\right]$$

(18)

which is independent of $B\phi$. Once $\kappa$ is beyond 1 as $A_r^*$ continuously increases, the surface friction coefficient would diverge from the one given by Eq. (18). In such case, surfaces with higher $B\phi$ yield larger friction coefficients. There is also an upper bond for a specified contact fraction. As $B\phi$ increasing, more and more asperities tend to deform elastically, and most normal force and shear force are contributed by these asperities. Ignore the second term in Eq. (9a) and the first and third term in Eq. (14a), and we can get the friction coefficient as

$$\mu_s / \alpha = f\left(-\left(\frac{2-D}{D}\right)^{1-D/2} \eta^* W^{*D-1} A_r^{*1-D/2}\right) \tag{19}$$

Eq. (19) together with Eq. (18) restricts $\mu_s/\alpha$ in the gray zone shown in Fig. 7.

The influences of surface topography parameters are displayed in Fig. 8 and Fig. 9. It is obvious that larger fractal dimension $D$ tends to cause a higher surface friction coefficient $\mu_s$. Usually, surfaces with larger $D$ will have larger root-mean-square roughness. From a micro perspective, with the increase in the fractal dimension $D$, the contribution of small sized spots to the friction coefficient will become more prominent, resulting in a larger $\mu_s$. However, the normalized roughness parameter $W^*$ plays an opposite effect. It can be inferred from Eq. (2) and Eq. (12) that a larger $W^*$ leads to increasing in the percentage of asperities of smaller radii, and decreasing of the asperity friction coefficient $\mu$. This decreasing in asperity friction coefficient eventually causes a lower surface friction coefficient $\mu_s$, as shown in Fig. 9.

Since the real contact area is difficult to measure, experiments tend to measure the curve of $\mu_s$ versus $P$. Fig. 10 displayed the variation of $\mu_s$ with respect to $P^*$ predicted by our model. As the normal load increase, the friction coefficient of rough surfaces $\mu_s$ transits from a constant value $\alpha$ to a power law dependence on the external load. Usually, the declining line is described by

$$\mu_s = CP^{-n} \tag{20}$$

where $C$ is a constant parameter independent of the loading process. It can be concluded from Fig. 10 that smaller $D$ leads to a larger exponent $n$, and larger $D$ results in a constant $\mu_s$ and a zero $n$. Consider $D$ approaching 1 or 2, and in both limiting cases $\kappa$ is far smaller than 1. Using Eq. (9b) and Eq. (14b), one has

$$\lim_{D \to 1} \mu_s = 3\alpha \left( B\phi\eta^{*-2} \right)^{1/3} P^{*-1/3}$$
$$\lim_{D \to 2} \mu_s = \alpha$$
(21)

Comparing Eq. (21) with Eq. (20), it can be obtained that $n \in (0, 1/3)$.

The friction coefficients of various materials have been measured experimentally [1, 3, 4], and the exponents were fitted as shown in Table 1. These exponents obtained from experiments are within the range of our prediction. With detailed information about surface geometry, this friction model can also predict the evolution of surface friction coefficient and thus is helpful to the design of friction pairs.

## 5. Conclusions

An unambiguous dependence of friction coefficient on the external load has been observed in previous rough surfaces friction experiments. Also, atomic-scale simulations have demonstrated a size dependence of single-asperity friction on contact area. In the present study, results from atomic-scale simulations are summarized and substituted into the conventional fractal contact model, and the formula of rough surface friction are conducted. The result shows that the friction coefficient is load dependent rather than a constant value. The surface friction coefficient transits from the same value as the local maximal friction coefficient to a power law dependence on both the real contact area and the normal load. The exponent of the power law relation between friction coefficient and normal load is predicted to be within the range of 0 to 1/3, which shows a good agreement with experimental results. The present study can help improving the understanding of rough surfaces friction.

**Acknowledgements**

Supports from the National Natural Science Foundation of China (Grant No. 11525209) are acknowledged.

**References**

[1] Archard, J. F. (1957). Elastic Deformation and the Laws of Friction. Proceedings of the Royal Society A-Mathematical, Physical and Engineering Sciences, 243(1233), 190–205.

[2] Broniec, C., & Lenkiewicz, W. (1982). Static Friction Processes under Dynamic Loads and Vibration. Wear, 80(3), 261-271.

[3] Rabinowicz, E., & Kaymaram, F. (1991). On the Mechanism of Failure of Particulate Rigid Disks. Tribology Transactions, 34(4), 618-622.

[4] Etsion, I., & Amit, M. (1993). The Effect of Small Normal Loads on the Static Friction Coefficient for Very Smooth Surfaces. Journal of Tribology-Transactions of the ASME, 115(3), 406-410.

[5] Ben-David, O., & Fineberg, J. (2011). Static Friction Coefficient Is Not a Material Constant. Physical Review Letters, 106, 254301.

[6] Braun, O. M., Steenwyk, B., Warhadpande, A., & Persson, B. N. J. (2016). On the

Dependency of Friction on Load: Theory and Experiment. European Physics Letters, 113(5), 56002.

[7] Fortunato, G., Ciaravola, V., Furno, A., Scaraggi, M., Lorenz, B., & Persson, B. N. J. (2017). Dependency of Rubber Friction on Normal Force or Load: Theory and Experiment. Tire Science and Technology, 45(1), 25-54.

[8] Dangnan, F., Espejo, C., Liskiweicz, T., Gester, M., & Neville, A. (2020). Friction and Wear of Additive Manufactured Polymers in Dry Contact. Journal of Manufacturing Processes, 59, 238-247.

[9] Ozaki, S., Matsuura, T., & Maegawa, S. (2020). Rate-, State-, and Pressure-Dependent Friction Model Based on the Elastoplastic Theory. Friction, 8(4), 768-783.

[10] Greenwood, J. A., & Williamson, J. B. P. (1966). Contact of Nominally Flat Surfaces. Proceedings of the Royal Society A-Mathematical, Physical and Engineering Sciences, 295(1442), 300–319.

[11] Bush, A. W., Gibson, R. D., & Thomas, T. R. (1975). The Elastic Contact of a Rough Surface. Wear, 35(1), 87–111.

[12] Song, H., Van der Giessen, E., & Liu, X. (2016). Strain Gradient Plasticity Analysis of Elasto-Plastic Contact Between Rough Surfaces. Journal of the Mechanics and Physics of Solids, 96, 18-28.

[13] Wang, G. F., Long, J. M. & Feng, X. Q. (2015). A Self-Consistent Model for the Elastic Contact of Rough Surfaces. Acta Mechanica, 226(2), 285-293.

[14] Mandelbrot, B. B., Passoja, D. E., & Paullay, A. J. (1984). Fractal Character of

Fracture Surfaces of Metals. Nature, 308(5961), 721–722.

[15] Majumdar, A., & Tien, C. L. (1990). Fractal Characterization and Simulation of Rough Surfaces. Wear, 136(2), 313–327.

[16] Majumdar, A., & Bhushan, B. (1991). Fractal Model of Elastic-Plastic Contact Between Rough Surfaces. Journal of Tribology, 113(1), 1-11.

[17] Yan, W., & Komvopoulos, K. (1998). Contact Analysis of Elastic-Plastic Fractal Surfaces. Journal of Applied Physics, 84(7), 3617–3624.

[18] Long, J. M., Wang, G. F., Feng, X. Q., & Yu, S. W. (2014). Influence of Surface Tension on Fractal Contact Model. Journal of Applied Physics, 115(12), 123522.

[19] Persson, B. N. J. (2001). Theory of Rubber Friction and Contact Mechanics. Journal of Chemical Physics, 115(8), 3840-3861.

[20] Persson, B. N. J. (2001). Elastoplastic Contact Between Randomly Rough Surfaces. Physical Review Letters, 87(11), 116101.

[21] Hamilton, G. M. (1983). Explicit Equations for the Stresses Beneath a Sliding Spherical Contact. Proceedings of the Institution of Mechanical Engineers Part C-Journal of Mechanical Engineering Science, 197(Mar), 53-59.

[22] Chang, W. R., Etsion, I., & Bogy, D. B. (1988). Adhesion Model for Metallic Rough Surfaces. Journal of Tribology-Transactions of the ASME, 110(1), 50-56.

[23] Chang, W. R., Etsion, I., & Bogy, D. B. (1988). Static Friction Coefficient Model for Metallic Rough Surfaces. Journal of Tribology-Transactions of the ASME, 110(1), 57-63.

[24] Kogut, L., & Etsion, I. (2003). A Semi-Analytical Solution for the Sliding

Inception of a Spherical Contact. Journal of Tribology-Transactions of the ASME, 125(3), 499-506.

[25] Kogut, L., & Etsion, I. (2004). A Static Friction Model for Elastic-Plastic Contacting Rough Surfaces. Journal of Tribology-Transactions of the ASME, 126(1), 34-40.

[26] Homola, A. M., Israelachvili, J. N., McGuiggan, P. M., & Gee, M. L. (1990). Fundamental Experimental Studies in Tribology: The Transition from Interfacial Friction of Undamaged Molecularly Smooth Surfaces to Normal Friction with Wear. Wear, 136(1), 65-83.

[27] Carpick, R. W., Agrait, N., Ogletree, D. F., & Salmeron, M. (1996). Measurement of Interfacial Shear (Friction) with an Ultrahigh Vacuum Atomic Force Microscope. Journal of Vacuum Science & Technology B, 14(2), 1289-1295.

[28] Li, Q. Y., & Kim, K. S. (2008). Micromechanics of Friction: Effects of Nanometre-Scale Roughness. Proceedings of the Royal Society A-Mathematical, Physical and Engineering Sciences, 464(2093), 1319-1343.

[29] Perčić, M., Zelenika, S., Mezić, I., Peter, R., & Krstulović, N. (2020). An Experimental Methodology for the Concurrent Characterization of Multiple Parameters Influencing Nanoscale Friction. Friction, 8(3), 577-593.

[30] Perčić, M., Zelenika, S., & Mezić, I. (2021). Artificial Intelligence-Based Predictive Model of Nanoscale Friction Using Experimental Data (in press). Friction.

[31] Hurtado, J. A., & Kim, K. S. (1999). Scale Effects in Friction of Single-Asperity

Contacts. Ⅰ. From Concurrent Slip to Single-Dislocation-Assisted Slip. Proceedings of the Royal Society A-Mathematical, Physical and Engineering Sciences, 455(1989), 3363-3384.

[32] Hurtado, J. A., & Kim, K. S. (1999). Scale Effects in Friction of Single-Asperity Contacts. Ⅱ. Multiple-Dislocation-Cooperated Slip. Proceedings of the Royal Society A-Mathematical, Physical and Engineering Sciences, 455(1989), 3385-3400.

[33] Sharp, T. A., Pastewka, L., & Robbins, M. O. (2016). Elasticity Limits Structural Superlubricity in Large Contacts. Physical Review B, 93(12), 121402.

[34] Sharp, T. A., Pastewka, L., Ligneres, V. L., & Robbins, M. O. (2017). Scale- and Load-Dependent Friction in Commensurate Sphere-On-Flat Contacts. Physical Review B, 96(15), 155436.

[35] Wang, J., Yuan, W. K., Bian, J. J., & Wang, G. F. (2020). A Semi-Analytical Model for the Scale-Dependent Friction of Nanosized Asperity. Journal of Physics Communications, 4(9), 095026.

[36] Gao, Y. F. (2010). A Peierls Perspective on Mechanisms of Atomic Friction. Journal of the Mechanics and Physics of Solids, 58(12), 2023-2032.

[37] He, G., Müser, M. H., & Robbins, M. O. (1999). Adsorbed Layers and the Origin of Static Friction. Science, 284(5420), 1650-1652.

[38] Peierls, R. (1940). The Size of a Dislocation. Proceedings of the Physical Society, 52, 34-37.

[39] Gourdon, D., & Israelachvili, J. N. (2003). Transitions Between Smooth and

Complex Stick-Slip Sliding of Surfaces. Physical Review E, 68(2), 021602.

[40] Mo, Y. F., Turner, K. T., & Szlufarska, I. (2009). Friction Laws at The Nanoscale. Nature, 457(7233), 1116-1119.

Table 1

The exponent $n$ of different friction pairs

| Friction pair | $n$ |
|---|---|
| Brass cylinders [1] | 0.333 |
| Barium titanate/magnetic disk | 0.082 |
| Al$_2$O$_3$-TiC/magnetic disk [3] | 0.082 |
| Al 2024/Ni | 0.102, 0.075 |
| Al 6061/Ni | 0.105, 0.203 |
| Al 7075/Ni [4] | 0.130, 0.165 |

**Figure captions**

Fig. 1. Dependence of the friction coefficient for repulsive sphere-on-flat friction on the contact size

Fig. 2. Variation of the normalized friction force $F^*$ with respect to the normalized real contact area $A_r^*$ for various $\eta^*$

Fig. 3. Variation of the normalized friction force $F^*$ with respect to the normalized real contact area $A_r^*$ for various $B\phi$

Fig. 4. Variation of the normalized friction force $F^*$ with respect to the normalized real contact area $A_r^*$ for various $D$

Fig. 5. Variation of the normalized external load $F^*$ with respect to the normalized real contact area $A_r^*$ for various $W^*$

Fig. 6. Variation of the normalized surface friction coefficient $\mu_s/\alpha$ with respect to the normalized real contact area $A_r^*$ for various $\eta^*$

Fig. 7. Variation of the normalized surface friction coefficient $\mu_s/\alpha$ with respect to the normalized real contact area $A_r^*$ for various $B\phi$

Fig. 8. Variation of the normalized surface friction coefficient $\mu_s/\alpha$ with respect to the normalized real contact area $A_r^*$ for various $D$

Fig. 9. Variation of the normalized surface friction coefficient $\mu_s/\alpha$ with respect to the normalized real contact area $A_r^*$ for various $W^*$

Fig. 10. Variation of the normalized surface friction coefficient $\mu_s/\alpha$ with respect to the normalized external load $P^*$ for various $D$ and $W^*$

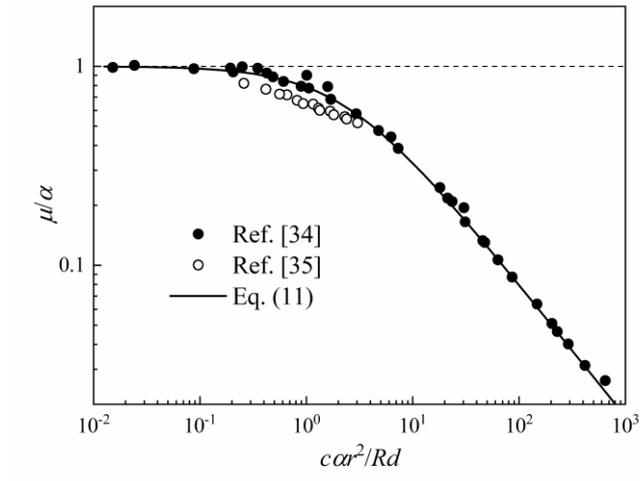

Figure 1

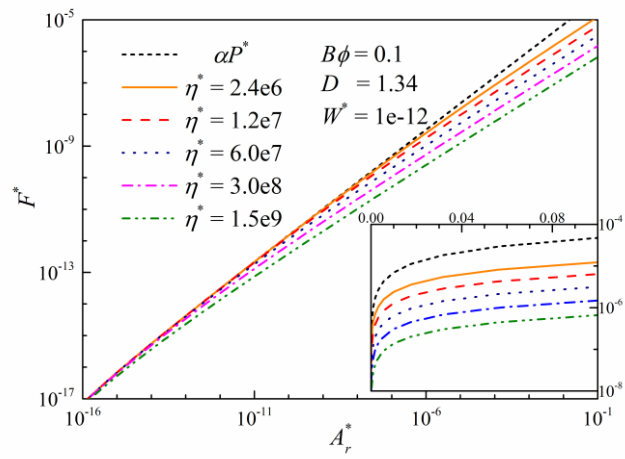

Figure 2

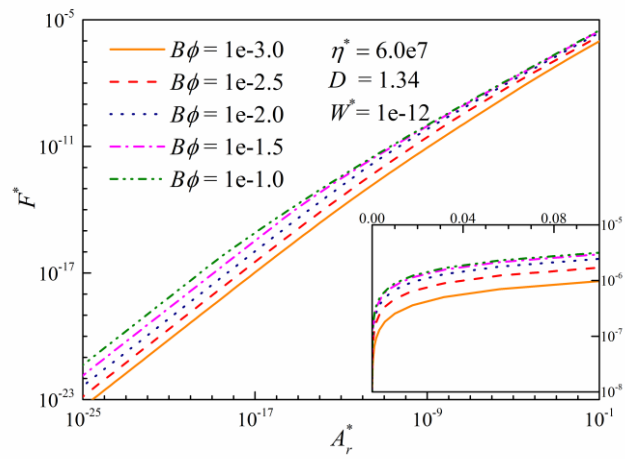

Figure 3

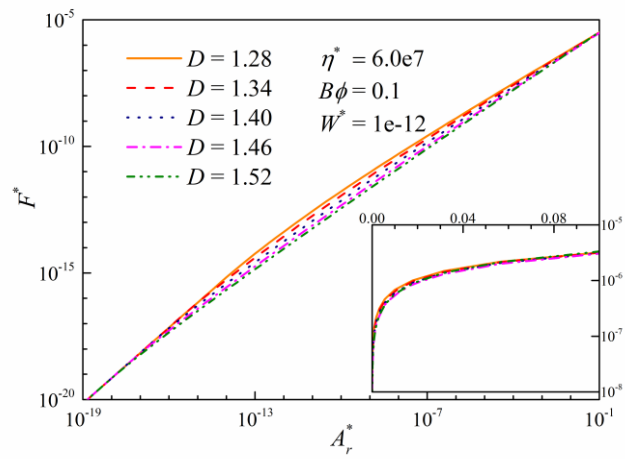

Figure 4

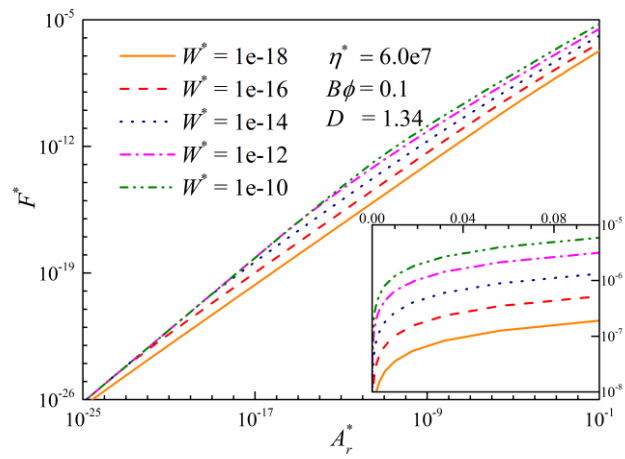

Figure 5

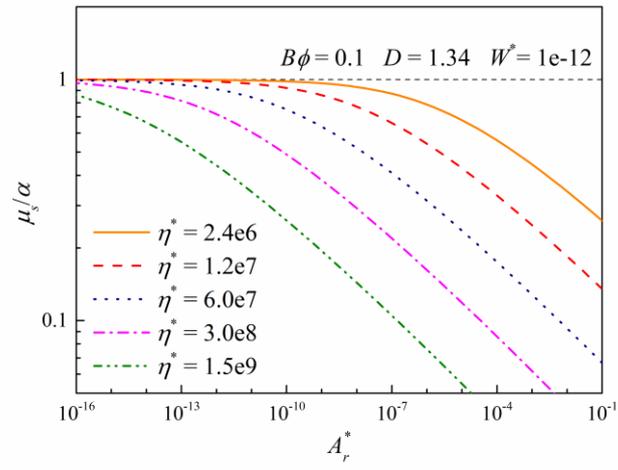

Figure 6

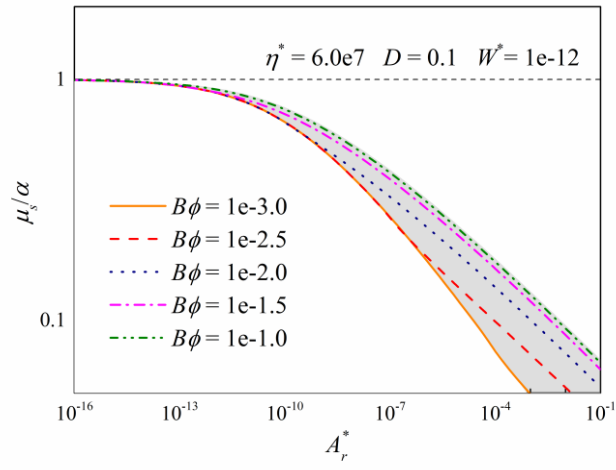

Figure 7

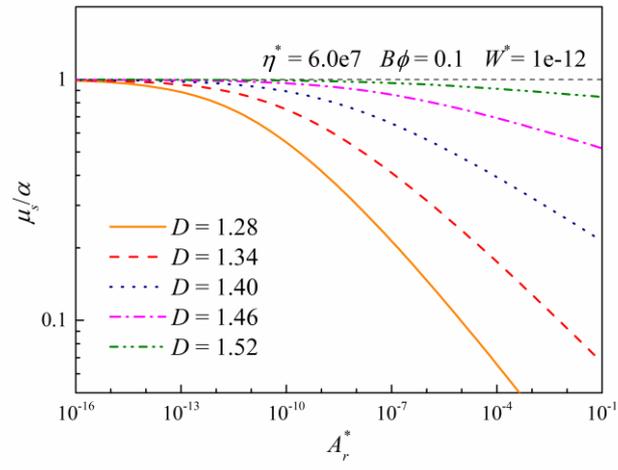

Figure 8

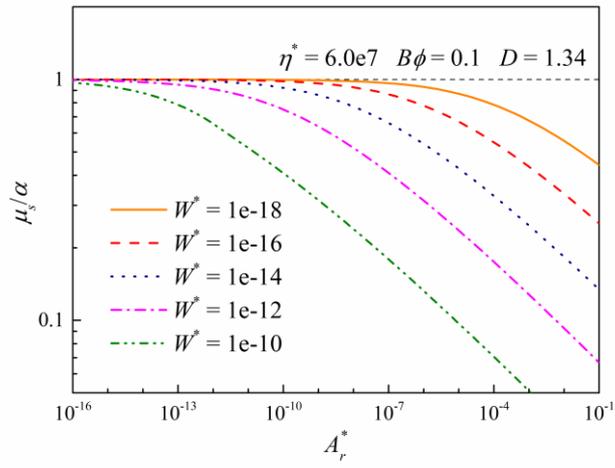

Figure 9

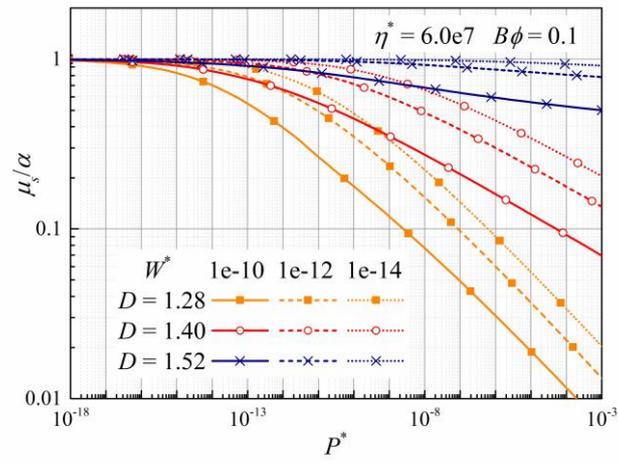

Figure 10